# Collaborative Learning for Annotation-Efficient Volumetric MR Image Segmentation

Yousuf Babiker M. Osman[1,2*], Cheng Li[1,3*], Weijian Huang[1,2], and Shanshan Wang[1,3#]

[1]Paul C. Lauterbur Research Center for Biomedical Imaging, Shenzhen Institute of Advanced Technology, Chinese Academy of Sciences, Shenzhen 518055, China

[2]University of Chinese Academy of Sciences, Beijing 100049, China

[3]Guangdong Provincial Key Laboratory of Artificial Intelligence in Medical Image Analysis and Application, Guangzhou 510080, China

[*]These authors contributed equally to this work

[#]Corresponding author: e-mail: sophiasswang@hotmail.com

## Abstract

**Background:** Deep learning has presented great potential in accurate MR image segmentation when enough labeled data are provided for network optimization. However, manually annotating 3D MR images is tedious and time-consuming, requiring experts with rich domain knowledge and experience.

**Purpose:** To build a deep learning method exploring sparse annotations, namely only a single 2D slice label for each 3D training MR image.

**Study Type:** Retrospective.

**Population:** 3D MR images of 150 subjects from two publicly available datasets were included. Among them, 50 (1,377 image slices) are for prostate segmentation. The other 100 (8,800 image slices) are for left atrium segmentation. Five-fold cross-validation experiments were carried out utilizing the first dataset. For the second dataset, 80 subjects were used for training and 20 were used for testing.

**Field Strength/Sequence:** 1.5 T and 3.0 T; axial T2-weighted and late gadolinium-enhanced, 3D respiratory navigated, inversion recovery prepared gradient echo pulse sequence.

**Assessment:** A collaborative learning method by integrating the strengths of semi-supervised and self-supervised learning schemes was developed. The method was trained using labeled central slices and unlabeled non-central slices. Segmentation performance on testing set was reported quantitatively and qualitatively.

**Statistical Tests:** Quantitative evaluation metrics including boundary intersection-over-union (B-IoU), Dice similarity coefficient, average symmetric surface distance, and relative absolute volume difference were calculated. Paired $t$-test was performed, and $p < 0.05$ was considered statistically significant.

**Results:** Compared to FS-LCS, MT, UA-MT, DCT-Seg, ICT, and AC-MT, the proposed method achieved a substantial improvement in segmentation accuracy, increasing the mean B-IoU significantly by more than 10.0% for prostate segmentation (proposed method B-IoU: 70.3%±7.6% vs. ICT B-IoU: 60.3%±11.2%) and by more than 6.0% for left atrium segmentation (proposed method B-IoU: 66.1%±6.8% vs. ICT B-IoU: 60.1%±7.1%).

**Data Conclusions:** A collaborative learning method trained using sparse annotations can segment prostate and left atrium with high accuracy.

**INTRODUCTION**

Accurate region of interest (ROI) segmentation in magnetic resonance images plays an important role in clinical practice.[1] For example, prostate cancer is among the most common cancers that impact men's healthcare.[2] Accurate prostate segmentation in MR images can help to localize prostate boundaries for radiotherapy and to perform prostate volume estimation for treatment outcome monitoring.[3] Similarly, cardiac diseases are the leading causes of morbidity and mortality worldwide.[4] Accurate segmentation of cardiac MR images, particularly the left atrium, is essential as it can provide clinicians with valuable information for the diagnosis, monitoring, and treatment of various cardiac diseases.[5] However, manually segmenting 3D prostate MR images or 3D cardiac MR images is resource-intensive and time-consuming, requiring experts with rich domain knowledge and experience. Therefore, developing automatic segmentation algorithms is clinically valuable.

Recently, deep learning has presented great potential in accurate MR image segmentation.[6] Provided with enough labeled data for network optimization, supervised learning-based methods have achieved good performance on various MR image segmentation tasks.[7-9] Nevertheless, supervised models are limited by their need for a large amount of annotated data to achieve good performance,[10] whereas high-quality annotated MR image datasets are difficult to collect.[11] To reduce the burden of manual annotations, abundant semi-supervised learning-based methods approaches have been developed.[12,13] Generating pseudo labels for unlabeled data during deep learning model optimization is one of the most prevalent approaches.[14,15] Here, pseudo labels are commonly generated by applying the trained model to the unlabeled data and using the model's prediction as the temporary labels. These pseudo labels are then utilized to further train the model.[14,15] Semi-supervised learning methods have made great progress by learning discriminative patterns in unlabeled data through pseudo label generation. Nevertheless, the quality of network self-generated pseudo labels is critical for the final segmentation task, and a certain number of densely annotated training data (i.e., assigning a class number to each pixel in the 3D MR images) are necessary to prevent large errors in pseudo label generation.[16]

Learning with sparse annotations is another direction that also aims to achieve a trade-off between annotation effort and segmentation performance. In this filed, sparse annotated training samples (e.g., only one or several slices are annotated in a 3D MR image) instead of densely annotated training samples are provided. Similar to semi-supervised learning, generating pseudo labels for unlabeled data is a widely employed technique in sparse annotation learning. One approach used in this context is analogous to semi-supervised learning methods, where the predictions made by the trained models are treated as pseudo labels and incorporated into the subsequent model training process.[17] Another approach involves employing self-supervised image registration to align neighboring image slices. Consequently, the annotations assigned to the labeled slices can be propagated to the unlabeled slices, leveraging the alignment information.[18,19] Although these sparse annotation learning works have validated the possibility of achieving promising results while requiring much less annotation effort, existing studies need either to select representative slices in a 3D image to annotate or to develop special modules to alleviate error accumulation during mask propagation in inference time.[17,19] The accuracy and efficiency still can be improved.

The objective of this study was to develop a deep learning method that can automatically segment MR images without too much human's manual labelling effort. The method explores the strength of both semi-supervised and self-supervised learning for collaboratively training the segmentation model. Specifically, the method tends to get accurate segmentation of the prostate and left atrium in 3D MR images by exploiting 2D slice labels for each 3D training MR image.

## MATERIALS AND METHODS

### Datasets

Datasets from two publicly available MR image segmentation challenges, involving prostate and left atrium segmentation, were utilized to evaluate the effectiveness of the proposed approach. Ethical permission for the use of data in this study should be used according to the challenges' requirements.[3,5]

The first dataset is the PROMISE12 dataset from the MICCAI prostate MR image segmentation challenge.[3] PROMISE12 provides 50 3D (one 3D image contains 15−54 slices) axial T2-weighted MR images of the prostate along with their corresponding segmentation ground-truth labels. The data were collected from four hospitals using various equipment and protocols: RUNMC (field strength: 3.0 T; resolution: 0.5–0.75/3.6–4.0 mm; scanner: Siemens), UCL (field strength: 1.5 T and 3T; resolution: 0.325–0.625/3–3.6 mm; scanner: Siemens), BIDMC (field strength: 3.0 T; resolution: 0.25/2.2–3 mm; scanner: GE), and HK (field strength: 1.5 T; resolution: 0.625/3.6 mm; scanner: Siemens).[3] In this study, five-fold cross-validation experiments were carried out using this dataset, and corresponding results are reported.

The second dataset is a left atrium dataset from the 2018 Atrial Segmentation Challenge.[5] A total of 100 3D late gadolinium-enhanced (LGE) MR scans with left atrium segmentation masks were included. The data were acquired with a 1.5 T Avanto scanner or a 3.0 T Verio whole-body scanner (Siemens).[5] The spatial resolution of each 3D LGE-MRI scan is $0.625 \times 0.625 \times 0.625$ mm³ with spatial dimensions of either 576 × 576 × 88 or 640 × 640 × 88 pixels. Each 3D MR scan has 88 slices on the z-axis. Following the previous works,[12,20] the same preprocessing procedures and data division strategy were adopted, where the provided 100 MR scans were divided into 80 scans for training and 20 scans for testing.

### *Proposed Collaborative Learning Method*

Let $D = \{D_1, D_2, D_3, \ldots, D_M\}$ represent the 3D training images ($M$ is the total number of 3D training images), each of which contains $N$ slices ($N$ could be different for different images) $D_m = \{D_m^1, D_m^2, D_m^3, \ldots, D_m^N\}$, where $D_m^n \in \mathbb{R}^{H \times W}$ represents one slice in the 3D image $D_m$ ($H$ and $W$ represent the height and width of the image). Under the sparse annotation learning setting, only the central slice of each 3D training image was labeled by the experts. Accordingly, the set of ground-truth labels that were available for model training was $Y = \{Y_1^c, Y_2^c, Y_3^c, \ldots, Y_M^c\}$ ($c$ refers to central slice) where $Y_m^c \in \mathbb{R}^{H \times W}$ has the same size as its corresponding central slice $D_m^c$. Figure 1 depicts the overview of the proposed collaborative learning method, which consists of three parts: pseudo label generation, pseudo label boosting, and target segmentation network learning.

***Figure 1 goes here***

### *Pseudo Label Generation by Semi-Supervised Learning*

To generate pseudo labels by semi-supervised learning, a segmentation network was firstly trained utilizing $\{D_m^c, Y_m^c\}$ ($m = 1, \ldots, M$) for $K_1$ epochs ($K_1$ will be explained in Eq. 3). Then, the unlabeled slices were introduced, and pseudo labels were generated and progressively finetuned for subsequent model training. For each unlabeled slice $D_m^n$ ($n \neq c$ to exclude the labeled central slice), it was input into the network and the category with the highest predicted probability was treated as the pseudo labels for each pixel:

$$\bar{Y}_m^n[i,j] = argmax_k \, f(D_m^n)[i,j,k] \tag{1}$$

Where $\bar{Y}_m^n[i,j]$ represents the pseudo label for the pixel [i,j] in $D_m^n$. $k$ refers to the different channels in the network prediction ($k \in \{0,1\}$).

The loss function adopted was the combination of Dice loss and cross-entropy loss. The same loss function was calculated for the labeled slices and unlabeled slices. To differentiate the contributions of manual labels and network self-generated pseudo labels, a dynamic weight parameter α(t) was introduced. The overall loss function was:

$$L_{SEMI} = [L_{Dice}(y,f) + \gamma . L_{CE}(y,f)] + \alpha(t) . [L_{Dice}(\acute{y},\acute{f}) + \gamma . L_{CE}(\acute{y},\acute{f})]$$

$$L_{Dice}(a,b) = 1 - \frac{2 . \sum_{i=1}^{\Omega} a_i . b_i + \sigma}{\sum_{i=1}^{\Omega} a_i + \sum_{i=1}^{\Omega} b_i + \sigma} \tag{2}$$

$$L_{CE}(a,b) = \frac{1}{\Omega}\left(\sum_{i=1}^{\Omega} a_i \log(b_i) - (1 - a_i) \log(1 - b_i)\right)$$

Where $f$ and $y$ represent the model's prediction and the ground-truth segmentation mask of the labeled slices. $\acute{y}$ and $\acute{f}$ represents the model's prediction and the pseudo label of unlabeled slices. $\Omega$ stands for the total number of pixels in the image. $\gamma$ is a constant used to balance the contributions of the cross-entropy and the Dice loss terms. $\sigma$ is a positive constant ($\sigma$=1) included to ensure numerical stability. $\alpha(t)$ is a dynamic coefficient designed to control the trade-off between the labeled and unlabeled slices:

$$\alpha(t) = \begin{cases} 0 \, , & t < K_1 \\ \frac{t-K_1}{K_2-K_1}\alpha_f, & K_1 \leq t \leq K_2 \\ \alpha_f, & K_2 \leq t \end{cases} \tag{3}$$

Where $t$ refers to the current training epoch number. $K_1$, $K_2$, and $\alpha_f$ are three constants determined empirically.

*Pseudo Label Generation by Self-Supervised Learning*

Following the previous work,[21] a self-supervised deformable registration model was employed. Briefly, during training, a registration model ($D_\theta$) was learned to generate the registration fields ($\Phi$) between neighboring slices. These registration fields were utilized to warp the images (deform the images) to accomplish the self-supervised image registration process. After the model training, these generated registration fields were then employed to propagate the labels from the labeled central slice to the unlabeled non-central slices to achieve the goal of pseudo label generation for the entire 3D MR images in the training set.

The label propagation began by treating the labeled central slice as the moving slice. Given the central slice $S_0$ and its adjacent slice $S_1$ in one direction ($S_{-1}$ in the other direction was treated in the same way), a pair of slices ($S_0$, $S_1$) was obtained, and $D_\theta$ was applied to estimate the spatial transformation ($\Phi_0$) that maps $S_0$ to $S_1$. Accordingly, the segmentation mask of $S_0$ ($y_0$) was transferred to $S_1$ by going through the same deformable transformations $y_1 = y_0 \circ \Phi_0$, where $\circ$ indicates the warp operation. Pseudo labels of all the unlabeled slices were generated by propagating the label from the central slice to the boarder slices iteratively.

*Boosting the Quality of Generated Pseudo Labels*

The pseudo labels generated by the semi-supervised and self-supervised learning methods were fused by considering the consistency between the two sets of pseudo labels. In practice, the Boolean set operations were utilized to measure the consistency. Consistent pseudo labels were defined as the intersection of the outputs of the two pseudo label generation methods:

$$y_{consis} = y_{SEMI} \cap y_{SSL} \qquad (4)$$

Where $y_{SEMI}$ and $y_{SSL}$ were the corresponding pseudo labels generated by semi-supervised learning and self-supervised learning, respectively. Inconsistent pseudo labels were those labels predicted to be positive by only one of the two methods. They were calculated by subtracting the intersection from the union of the two types of pseudo labels:

$$y_{inconsis} = (y_{SEMI} \cup y_{SSL}) - (y_{SEMI} \cap y_{SSL}) \qquad (5)$$

*Optimizing the Target Network for Volumetric Medical Image Segmentation*

Finally, to segment the target objects in volumetric MR images, an encoder-decoder segmentation network was trained in a fully supervised manner utilizing either the provided manual labels or the generated pseudo labels. The segmentation loss function utilized to train the network was combined of Dice loss and cross-entropy loss:

$$L_{Seg\_Certain} = L_{Dice}(Y_{certain}, F) + \gamma_{certain}.L_{CE}(Y_{certain}, F)$$

$$Y_{certain} = \begin{cases} y, & X \in \{S_i^c\} \\ y_{consis}, & X \in \{S \setminus S_i^c\} \end{cases} \qquad (6)$$

$$L_{Seg\_Uncertain} = L_{Dice}(Y_{uncertain}, F) + \gamma_{uncertain}.L_{CE}(Y_{uncertain}, F)$$

$$Y_{uncertain} = y_{inconsis}, \quad X \in \{S \setminus S_i^c\}$$

Where $F$ was the prediction of the model. $L_{Seg\_Certain}$ represents the loss function calculated for the certain labels, which include the ground-truth manual labels and the consistent pseudo labels. $L_{Seg\_Uncertain}$ refers to the loss function calculated for the uncertain labels, which include the inconsistent pseudo labels. $X$ was the input image. $y$ refers to the provided manual labels. $S_i^c$ represents the central slice set, and $S \setminus S_i^c$ indicates the unlabeled non-central image slice set. $\gamma_{certain}$ and $\gamma_{uncertain}$ were constants to control the contributions of the Dice loss and cross-entropy loss. In this work, $\gamma_{certain}$ and $\gamma_{uncertain}$ were set to 1.0 and 0.1, respectively. The certain and uncertain labels were treated differently by assigning different loss weights:

$$L_{Total\_Seg} = L_{Seg\_Certain} + \beta \cdot L_{Seg\_Uncertain} \qquad (7)$$

Here, $\beta$ was a constant to control the different contributions of the certain and uncertain labels. $\beta$ was set to 0.5 in implementation to properly decrease the possible negative effects of the uncertain labels (inconsistent pseudo labels).

### Network Architectures

A U-shaped network was adopted in the semi-supervised learning method.[9] The architecture comprised a contracting path for capturing context and a symmetric expanding path for precise localization. In the contracting path, the kernel number was doubled at each down-sampling step. Each step in the expanding path consisted of a feature map up-sampling module, reducing the number of features by half, followed by concatenation with the corresponding feature maps from the contracting path. Different from the original model,[9] in this study, zero padding was incorporated to keep the feature resolutions unchanged in the convolution layers. Besides, a batch normalization layer was added after each convolution layer to facilitate training and enhance performance. The same segmentation network was also employed to generate the deformable transformation fields and perform the final target object segmentation. Nevertheless, it should be noted that the proposed method is flexible in terms of the network architecture, and any segmentation network can be employed.

### Training Strategies

The semi-supervised learning method followed the following specific network training schedule: Step 1 – Train the network with only the labeled slices for 50 epochs in a supervised learning manner (i.e., $K_1$=50 in Eq. 3) and save the network model's weights. The Adam (Adaptive Moment Estimation) optimizer was employed,[22] with a learning rate of 0.0001. Step 2 – Gradually incorporate the information of the unlabeled slices. The model saved in the first step was used to generate pseudo labels for unlabeled images slices. These pseudo labels were recalculated after every weight update training epoch. The relevant hyper-parameters in Eq. 3 were set as $\alpha_f = 3$ and $K_2 = 100$.

For the deformable transformation network $D_\theta$, the same Adam optimizer was utilized. The batch size was set to 32 and the learning rate was set to 0.01.

The training of the target network for final volumetric MR image segmentation still employed the Adam optimizer. The learning rate was set to 0.0001 and decayed by half every 30 epochs, with a batch size of 4, for a total of 100 training epochs. Data augmentation techniques including rotation and flipping were used to boost the use of the limited training data.

All experiments were conducted utilizing a NVIDIA GeForce RTX 2080 Ti GPU. PyTorch was utilized to implement the deep learning approach.[29]

### *Comparison Methods*

The results of the proposed method were compared with five state-of-the-art semi-supervised learning approaches, including mean teacher (MT),[23] uncertainty-aware mean teacher (UA-MT),[24] deep co-training (DCT-Seg),[25] interpolation consistency training (ICT),[26] and ambiguity-consensus mean-teacher (AC-MT).[20] One lower bound baseline, fully supervised training with only the labeled central slice (FS-LCS), and one upper bound baseline, fully supervised training with all the training images (FS),[9] were also included.

For fair comparisons, all semi-supervised learning comparison methods were trained using the same settings as the proposed method. Publicly available codes of these methods were utilized to implement them.

*Experimental Setups: Feature Importance Analysis*

Four sets of feature importance analysis experiments were conducted to fully validate the effectiveness of the proposed method. Set 1: The importance of employing the two pseudo label generation methods were investigated. Specifically, experiments with either only the semi-supervised pseudo label generation method (Semi-PL-Net) or only the self-supervised pseudo label generation method (Self-PL-Net) were performed. Set 2: The sensitivity of segmentation performance on the hyper-parameter $\alpha_f$ in Eq.3 were evaluated. Specifically, different values of $\alpha_f$ (i.e., 0.1, 0.5, 1.0, 3.0, 5.0, 7.0, and 9.0) were experimented with using Semi-PL-Net. Set 3: Experiments with the baseline FS-LCS were conducted to explore how many slices need to be annotated in each 3D training MR image for FS-LCS to achieve a similar segmentation performance as the proposed method trained with one labeled slice per 3D training image. Specifically, different numbers of annotated slices (i.e., 2 slices, 3 slices, 5 slices, 6 slices, and 7 slices) were provided, and FS-LCS was trained and evaluated. Set 4: Different pseudo label fusion methods, intersection, union, and the proposed consistency-based fusion, were investigated. All feature importance analysis experiments were performed using the PROMISE12 dataset.

*Statistical Analysis*

For quantitative evaluation, boundary intersection-over-union (B-IoU), Dice similarity coefficient (DSC), average symmetric surface distance (ASSD), and relative absolute volume difference (RAVD) were calculated. The mean values and standard deviations of the metrics were calculated and reported based on the test data. Higher B-IoU and DSC, as well as lower ASSD and RAVD, indicate better segmentation results.

$$\text{B-IoU} = \frac{|B_{\hat{y}} \cap B_y|}{|B_{\hat{y}} \cup B_y|} \tag{8}$$

$$DSC = \frac{2TP}{2TP + FP + FN} \qquad (9)$$

$$ASSD = \frac{1}{|B_{\hat{y}}| + |B_y|} \left( \sum_{x \in B_{\hat{y}}} d(x, B_y) + \sum_{x \in B_y} d(x, B_{\hat{y}}) \right) \qquad (10)$$

$$RAVD = abs(\frac{FP - FN}{TP + FN}) \qquad (11)$$

Where TP, FP, and FN stand for true positive, false positive, and false negative predictions, respectively. $B_{\hat{y}}$ and $B_y$ represent the predicted and ground-truth segmentation boundary points. $d(.,.)$ represents the Euclidean distance between the corresponding point on one surface and its nearest neighbor on the second surface. $|\cdot|$ indicates non-zero number count. $abs(\cdot)$ represents absolute value.

Additionally, employing two-tailed paired $t$-test along with a significance threshold of $p < 0.05$, statistical significance tests were performed on all evaluation metric values to compare the segmentation performance of the proposed method to the comparison approaches.

## RESULTS

### *Results on Prostate Segmentation*

The quantitative results of different methods on prostate segmentation are listed in Table 1. Example visualization results of different methods are plotted in Figure 2. The lower bound method, FS-LCS, achieved the lowest B-IoU of 48.6%±16.9% and DSC of 63.4%±18.6%, as well as the largest RAVD of 0.50±0.44, with medium ASSD of 5.08 mm±3.24 mm. The fully supervised upper bound method, FS, obtained the highest B-IoU and DSC scores of 78.3%±6.5% and 87.6%±4.1% and the lowest ASSD and RAVD scores of 1.37 mm±0.51 mm and 0.13±0.12. The five state-of-the-art semi-supervised approaches (MT, ICT, DCT-Seg, UA-MT, and AC-MT) obtained increased B-IoU and DSC scores with decreased RAVD when compared to FS-LCS. Among those semi-supervised methods, ICT generated the best B-IoU, DSC, and ASSD scores of 60.3%±11.2%, 74.6%±9.2%, and 4.92 mm±5.38 mm, respectively. Compared to ICT, the proposed method (Ours) increased the B-IoU and DSC significantly by 10.0% and 7.5%, respectively, and decreased the ASSD and RAVD significantly by 2.72 mm and 0.07, respectively.

***Table 1 goes here***

***Figure 2 goes here***

### *Results on Left Atrium Segmentation*

The quantitative results of different methods on left atrium segmentation are listed in Table 2. Example visualization results of different methods are plotted in Figure 3 and Figure 4. Here, similar to prostate segmentation, FS obtained the highest B-IoU and DSC scores of 75.1%±5.7% and 85.6%±3.8% and the lowest ASSD and RAVD scores of 1.04 mm±0.20 mm and 0.05±0.03. Compared to FS-LCS and the five existing semi-supervised approaches, the proposed method increased the B-IoU significantly by more than 6.0% (Ours: 66.1%±6.8% vs. ICT: 60.1%±7.1%) and decreased the ASSD and RAVD significantly by more than 1.62 mm (Ours: 2.95 mm±0.78 mm vs. FS-LCS: 4.57 mm±5.53 mm) and 0.07 (Ours: 0.15±0.12 vs. AC-MT: 0.22±0.17), respectively. For the DSC score, the proposed method showed no significant difference when compared to FC-LCS (p = 0.051), MT (p = 0.794), UA-MT (p = 0,403), and ICT (p = 0.720). In the meantime, it can be observed in Figure 3 that all methods generated very accurate left atrium segmentation maps in the central slices. However, according to Figure 4, the segmentation performance decayed obviously in top or bottom slices.

***Table 2 goes here***

***Figure 3 goes here***

***Figure 4 goes here***

### *Results from Feature Importance Analysis Experiments*

For Set 1 feature importance analysis experiments, Table 3 gives the quantitative results. Compared to FS-LCS, Semi-PL-Net increased the B-IoU and DSC scores by 11.6% and 11.0%. Self-PL-Net increases the values by 15.9% and 14.8%. By fusing the pseudo labels generated by the two methods, the final model increased the B-IoU and DSC scores significantly by more than 18% when compared to FS-LCS. Figure 5 plots several examples of the pseudo labels generated by different methods. Similar observation can be made that the proposed method generated more informative pseudo labels.

***Table 3 goes here***

***Figure 5 goes here***

The DSC scores obtained from Set 2 feature importance analysis experiments with different $\alpha_f$ values are plotted in Figure 6. Overall, the segmentation performance shows a first increase and then decrease pattern, and $\alpha_f = 3$ achieved the best DSC score of 74.4%±9.8%. $\alpha_f = 3$ was utilized for all other relevant experiments.

***Figure 6 goes here***

The quantitative segmentation results of Set 3 feature importance analysis experiments are listed in Table 4. With the increasing of labeled slices, the segmentation accuracy was also increased. For FS-LCS, it needed at least 6 annotated slices for each 3D training image to achieve comparable segmentation results (B-IoU of 70.2%±13.1% and DSC of 81.7%±10.4%) with those that generated by the proposed method trained with 1 annotated slice for each 3D training image (B-IoU of 70.3%±7.6% and DSC of 82.1%±5.6%).

***Table 4 goes here***

Finally, for Set 4 feature importance analysis experiments, results are plotted in Figure 7. Among the three methods, pseudo label fusion by union operation generated the lowest B-IoU of 55.9%±11.4% and DSC of 71.0%±10.1%. Fusion by intersection gave better performance with B-IoU of 68.6%±8.9% and DSC of 81.0%±6.8%. Although the enhancement was not significant when compared to intersection ($p = 0.380$ for B-IoU and $p = 0.337$ for DSC), the proposed fusion method achieved the best scores with the highest B-IoU of 69.7%±7.6% and DSC of 81.9%±5.6%.

***Figure 7 goes here***

## DISCUSSION

Many efforts have been made to address various challenges in the field of supervised deep learning for MR imaging.[6] However, the scarcity of annotated data remains to be a major limiting factor.[10] For example, a 3D MR scan typically comprises tens to hundreds of slices,[3,5] and experts need to go through all these slices one by one to delineate the target regions, which is a very time-consuming process that can take several hours to accomplish.[17,27] The development of new techniques with reduced annotation requirements holds great potential for real-world clinical applications. In this study, a collaborative learning method for volumetric MR image segmentation via exploring sparse annotations, namely only single 2D slice label for each 3D training MR image, was proposed. This approach has a high potential to reduce the annotation burden by utilizing a large amount of unannotated samples. The key idea behind this collaborative learning method was to integrate the strengths of semi-supervised and self-supervised learning schemes, enabling the generation of more informative pseudo labels for unlabeled data while jointly learning from both inter-slice and intra-slice knowledge.

The effectiveness of the proposed method has been validated through extensive experiments on two publicly available MR image datasets for prostate and left atrium segmentation. Both quantitative and qualitative results were reported. Among the four quantitative evaluation metrics, B-IoU provides a measure of the overall overlap between boundaries of the predicted and ground truth masks and is less affected by variations in mask size. DSC focuses on the average size of the two masks. RAVD measures the relative volume difference between the predicted and ground truth masks, which is less affected by the boundary difference. ASSD, on the other hand, is a boundary-based metrics, which is less affected by the volume difference. These four metrics were calculated to comprehensively evaluate the segmentation performance. Experimental results showed that semi-supervised learning could obtain better results than the baseline FS-LCS method, demonstrating the important information unlabeled data can provide for deep learning model training. Compared to existing state-of-the-art semi-supervised approaches, the proposed collaborative learning method achieved enhanced segmentation performance on both datasets, validating its effectiveness on handling annotation-limited MR image segmentation tasks. Besides, although the training of the proposed method took several hours, it took less than one minute to generate the segmentation masks for a whole 3D MR image during inference. Nevertheless, the performance of the proposed method was still worse than the fully supervised method, FS, indicating the reliance of deep learning models on clinical knowledge input. Feature importance analysis studies validated the importance of each component to the proposed method. Moreover, it was found that FS-LCS required a minimum of 6 annotated slices per 3D MR image to achieve comparable accuracy with the proposed method. This observation indirectly suggested that the proposed approach can effectively leverage unlabeled image data, resulting in approximately 83.3% (5/6) reduction in annotation effort compared to FS-LCS.

The proposed collaborative learning method is believed to be well-suited for MR imaging. There are three reasons. Firstly, MR imaging offers excellent soft tissue contrast and high 3D spatial resolution. These properties are crucial for accurate image segmentation, making MR imaging a suitable target modality for the proposed method. Moreover, the high spatial resolution is particularly important for the success of the self-supervised learning-based pseudo label generation component in the proposed collaborative learning method, which relied on precise registration between adjacent slices that can be facilitated by the high spatial resolution of MR images. Due to the same reason, compared to the different MR image contrasts, the proposed method is more limited by the image resolution. Secondly, MR imaging is widely considered as the gold standard imaging modality in clinic practice. It can be utilized to image different body parts without exposing patients to ionizing radiation. By providing a method for the segmentation of MR images, the proposed method aimed to gain valuable insights into its potential for real-world clinical applications. Last but not least, there are many publicly available MR image datasets that can be used for segmentation-related studies. These datasets provide a valuable resource for training and evaluating the proposed method. In this study, experiments were conducted for prostate and left atrium segmentation as these two tasks are widely recognized as crucial in clinical practice. They encompass a diverse array of challenges that often arise in volumetric MR image segmentation. By addressing these challenges, we aimed to enhance the understanding and performance of segmentation techniques in the context of complex MR imaging data.

### *Limitations*

The proposed method has demonstrated promising results with the limited training annotations, which is particularly valuable for MR imaging where annotated data are expensive to obtain. However, it is important to note that expert annotations for the central slice of each 3D image were still required. The proposed method was indeed a semi-supervised learning method, which was also the reason for the selection of the comparison methods (all were semi-supervised learning approaches) in this study. To further relieve the reliance on manual annotations, unsupervised learning methods, including self-supervised learning approaches, should be investigated in the future.[28,29] Furthermore, it was observed that the segmentation performance decayed at the top and bottom slices of the left atrium. This performance discrepancy could be attributed to the different data distributions of these slices compared to the labeled central slices, which increased the difficulty of generating accurate pseudo labels and consequently affected the overall performance.[30] It is worth to investigate the method for more complex anatomy, such as the mitral valve and pulmonary veins.[5] Introducing more domain knowledge related to MR imaging and the segmentation task might be one possible solution to this issue. Finally, in this study, only two datasets with relatively small sample sizes were experimented with in a retrospective manner. Validating the proposed method utilizing larger and multi-institutional datasets prospectively will be more clinically relevant.

## *Conclusions*

This study proposes a pseudo label collaborative learning method for volumetric MR image segmentation with sparse annotations, for which the ground-truth labels are provided for just one 2D slice of each 3D training image. To make full use of the unlabeled image slices, pseudo labels were generated by two different methods, a semi-supervised learning method and a self-supervised learning method. Furthermore, pseudo label fusion based on consistency was conducted to boost the quality of network self-generated labels, and thus, enhanced final segmentation performance was achieved. Extensive experiments on two publicly available datasets validated the effectiveness of the proposed method when compared to existing methods. In real-world clinical applications, the proposed method has a potential to offer a good solution when annotating large quantities of 3D MR images for model learning is infeasible.

## ACKNOWLEDGEMENTS

This research was partly supported by Scientific and Technical Innovation 2030-"New Generation Artificial Intelligence" Project (2020AAA0104100, 2020AAA0104105), the National Natural Science Foundation of China (62222118, U22A2040), Guangdong Provincial Key Laboratory of

**TABLE 1. Quantitative Results (Mean±S.D.) on PROMISE12 Dataset by 5-Fold Cross-Validation Experiments**

| Methods | B-IoU [%] | DSC [%] | ASSD [mm] | RAVD |
|---|---|---|---|---|
| FS-LCS | 48.6*±16.9 | 63.4*±18.6 | 5.08*±3.24 | 0.50*±0.44 |
| MT[23] | 58.9*±13.4 | 73.1*±11.6 | 5.76*±5.40 | 0.28*±0.26 |
| UA-MT[24] | 59.5*±11.3 | 74.0*±9.5 | 5.36*±5.52 | 0.25*±0.19 |
| DCT-Seg[25] | 56.0*±13.5 | 70.8*±12.1 | 8.06*±7.95 | 0.45*±0.43 |
| ICT[26] | 60.3*±11.2 | 74.6*±9.2 | 4.92*±5.38 | 0.25*±0.20 |
| AC-MT[20] | 57.5*±14.4 | 71.0*±12.2 | 6.24*±5.89 | 0.29*±0.29 |
| **Ours** | **70.3±7.6** | **82.1±5.6** | **2.20±1.01** | **0.18±0.12** |
| **FS[9]** | **78.3±6.5** | **87.6±4.1** | **1.37±0.51** | **0.13±0.12** |

Data with * indicates $p < 0.05$ compared with the proposed method (Ours).

**TABLE 2. Quantitative Results (Mean±S.D.) on Left Atrium Dataset**

| Methods | B-IoU [%] | DSC [%] | ASSD [mm] | RAVD |
|---|---|---|---|---|
| FS-LCS | 55.7*±28.1 | 66.3±30.9 | 4.57*±5.53 | 0.37*±0.33 |
| MT[23] | 59.8*±6.9 | 78.0±5.7 | 5.64*±1.36 | 0.40*±0.26 |
| UA-MT[24] | 58.8*±7.9 | 78.0±6.8 | 6.57*±2.11 | 0.39*±0.27 |
| DCT-Seg[25] | 57.3*±9.1 | 75.3*±8.8 | 7.72*±2.84 | 0.24*±0.19 |
| ICT[26] | 60.1*±7.1 | 78.8±5.9 | 5.46*±1.64 | 0.38*±0.30 |
| AC-MT[20] | 53.8*±13.3 | 70.0*±14.1 | 9.17*±4.42 | 0.22*±0.17 |
| **Ours** | **66.1±6.8** | **79.4±5.1** | **2.95±0.78** | **0.15±0.12** |
| **FS[9]** | **75.1±5.7** | **85.6±3.8** | **1.04±0.20** | **0.05±0.03** |

Data with * indicates $p < 0.05$ compared with the proposed method (Ours).

**TABLE 3. Quantitative Results (Mean±S.D.) of Set 1 Feature Importance Analysis Experiments on Prostate Segmentation**

| Methods | B-IoU [%] | DSC [%] | ASSD [mm] | RAVD |
|---|---|---|---|---|
| FS-LCS | 48.6*±16.9 | 63.4*±18.6 | 5.08*±3.24 | 0.50*±0.44 |
| Semi-PL-Net | 60.2*±12.3 | 74.4*±9.8 | 3.95*±1.75 | 0.52*±0.43 |
| Self-PL-Net | 64.5*±6.8 | 78.2*±5.2 | 2.87*±0.80 | 0.35*±0.22 |
| Ours (Semi-PL-Net & Self-PL-Net) | 70.3±7.6 | 82.1±5.6 | 2.20±1.01 | 0.18±0.12 |

Data with * indicates $p < 0.05$ compared with the proposed method (Ours).

**TABLE 4. Quantitative Results (Mean±S.D.) of Set 3 Feature Importance Analysis Experiments on Prostate Segmentation**

| Annotated slices per 3D MR image | B-IoU [%] | DSC [%] | ASSD [mm] | RAVD |
|---|---|---|---|---|
| 2 Slices | 60.1 ±17.3 | 73.3 ±16.7 | 3.65 ±4.79 | 0.30 ±0.37 |
| 3 Slices | 63.9 ±14.5 | 76.9 ±12.5 | 2.66 ±1.43 | 0.23 ±0.20 |
| 5 Slices | 69.5 ±13.4 | 80.9 ±11.6 | 2.14 ±1.28 | 0.20 ±0.18 |
| 6 Slices | 70.2 ±13.1 | 81.7 ±10.4 | 2.13 ±1.54 | 0.19 ±0.18 |
| 7 Slices | 72.8 ±8.9 | 83.9 ±6.5 | 1.86 ±1.00 | 0.17 ±0.14 |

## List of Figures

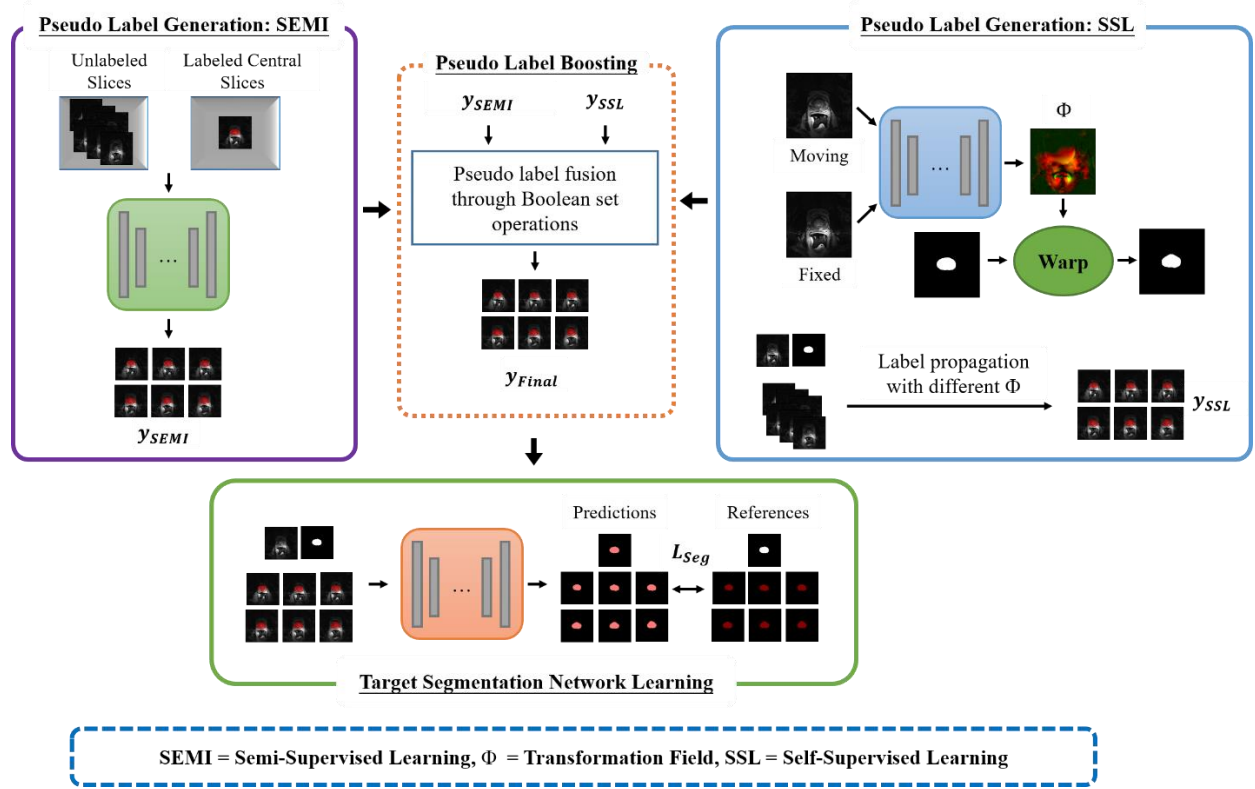

**FIGURE 1:** Overview of the proposed collaborative learning method, which contains three parts: pseudo label generation, pseudo label boosting, and target segmentation network learning.

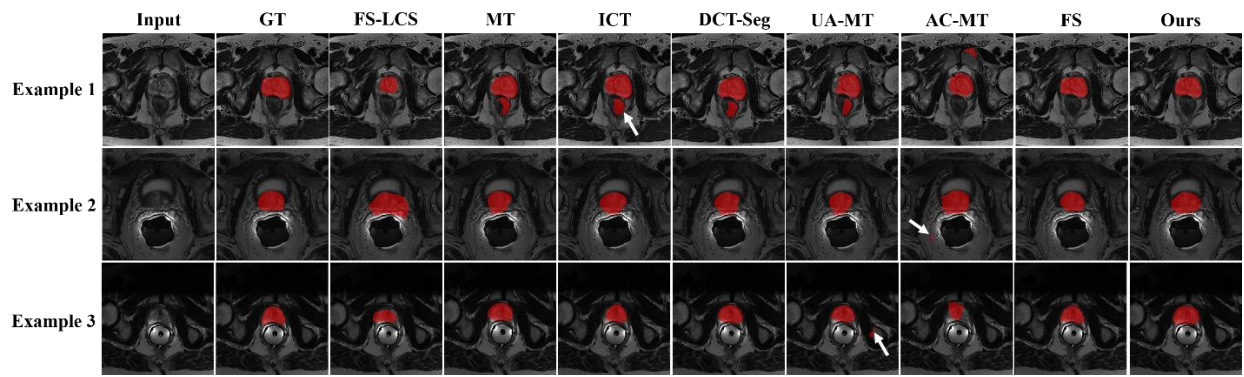

**FIGURE 2:** Examples prostate segmentation results of different methods. Red color indicates the segmented prostate segmentation regions.

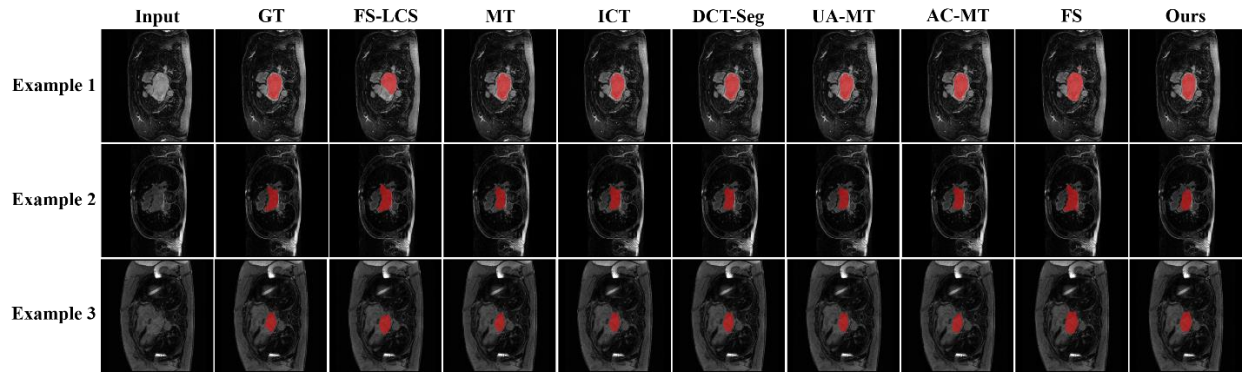

**FIGURE 3: Examples left atrium segmentation results of different methods. Red color indicates the segmented left atrium regions.**

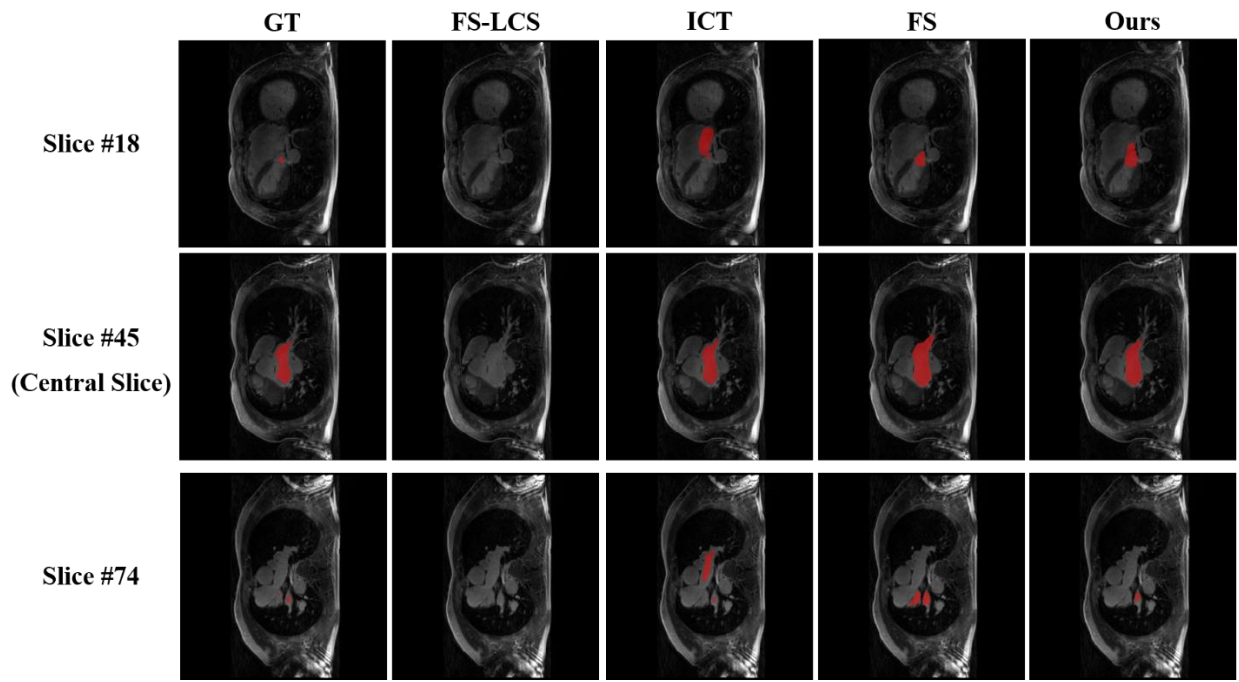

**FIGURE 4: Left atrium segmentation results of different methods in different slices of a 3D MR image. Red color indicates the segmented left atrium regions.**

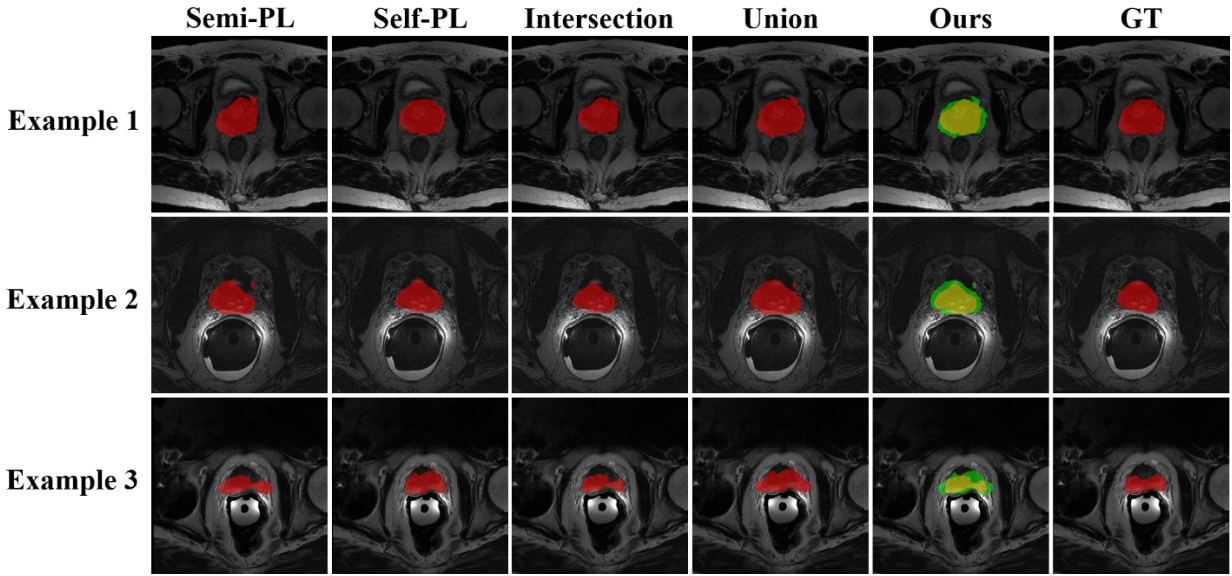

**FIGURE 5:** Example pseudo labels generated by different methods. Red color regions indicate the segmented prostate segmentation. Yellow and green colors indicate the consistent and inconsistent pseudo label regions obtained by the designed fusion method, respectively.

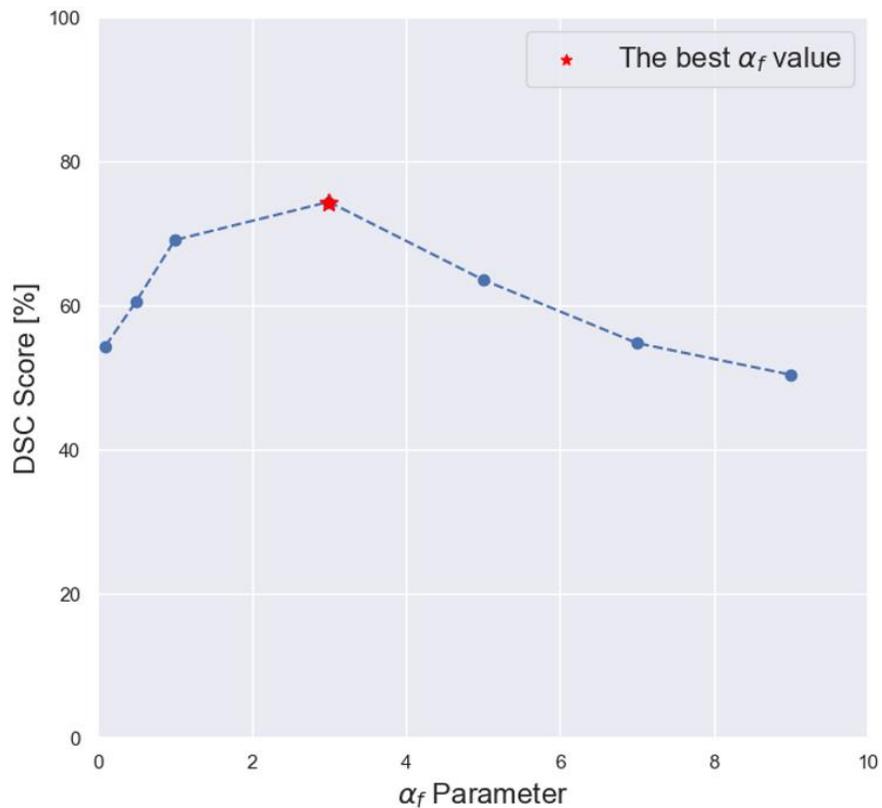

**FIGURE 6:** Prostate segmentation results of Semi-PL-Net with different $\alpha_f$ values in Eq. 3.

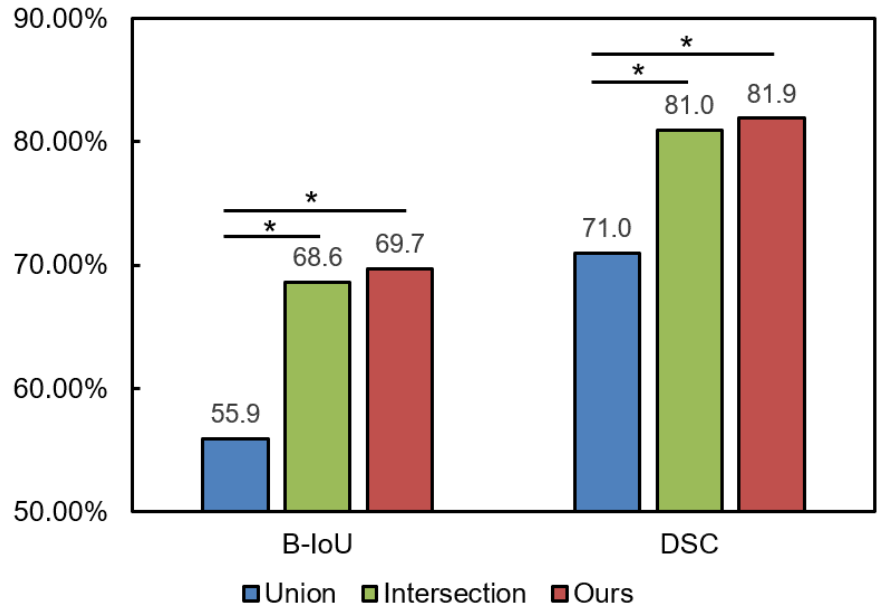

**FIGURE 7: Prostate segmentation results of the target segmentation network using different label fusion methods. * indicates $p < 0.05$ between the two results.**